\documentclass[12pt,preprint]{aastex}
\usepackage{natbib,graphics,graphicx,multirow,amsmath,cancel}
\usepackage{lscape}
\shorttitle{Coronal BP oscillation}
\shortauthors{K. Chandrashekhar, A. Sarkar}
\begin{document}
 \title{Modeling solar coronal bright point oscillations \\ with multiple nanoflare heated loops}
 \author{K Chandrashekhar, Aveek Sarkar}
 \affil{Center of Excellence in Space Sciences \\ Indian Institute of Science Education And Research Kolkata \\ Mohanpur -- 741246}
 \email{chandra@sdu.edu.cn, aveek.sarkar@iiserkol.ac.in}
\begin{abstract}
Intensity oscillations of coronal bright points (BPs) have been studied for past several years. It has been known for a while that these BPs are closed magnetic loop like structures. However, initiation of such intensity oscillations is still an enigma. There have been many suggestions to explain these oscillations, but modeling of such BPs have not been explored so far. Using a multithreaded nanoflare heated loop model we study the behavior of such BPs in this work. We compute typical loop lengths of BPs using potential field line extrapolation of available data~\citep{chandra13}, and set this as the length of our simulated loops. We produce intensity like observables through forward modeling and analyze the intensity time series using wavelet analysis, as was done by previous observers. The result reveals similar intensity oscillation periods reported in past observations. It is suggested these oscillations are actually shock wave propagations along the loop. We also show that if one considers different background subtractions, one can extract adiabatic standing modes from the intensity time series data as well, both from the observed and simulated data. 
\end{abstract}

\keywords{methods: data analysis, methods: numerical, Sun: corona, Sun: flares, Sun: oscillations, Sun: UV radiation}
\section{Introduction}
Coronal bright points (BPs) are ubiquitous features of the quiet-Sun and of the coronal-hole regions. These features were first sighted by~\cite{vaiana73} while observing the Sun using rocket born X-ray telescope. Later,~\cite{golub74} determined the lifetime of such X-ray BPs to be about eight hours using Skylab X-ray images. BPs are also observed in EUV~\citep{habbal81}. It was found~\citep{krieger71, golub74} that these BPs are actually small magnetic loop like structures associated with magnetic bipolar regions in the photosphere. The typical size of these BPs are about $10^8$ Km$^2$~\citep{golub74}.

The most distinctive feature of these BPs is their temporal variability during their lifetime~\citep{sheeley79, habbal81, abhishek11, chandra13}. It has been suggested that this temporal variability is either due to the interactions of the solar $p$-mode oscillations with the magnetic fields in the corona resulting into magnetosonic waves~\citep{bogdan03, kuridze08}, or due to recurrent magnetic reconnections~\citep{doyle06,tian08}. Our primary aim in this work is to investigate the extent to which the latter possibility contributes to the intensity oscillations of the BPs.

Since BPs are composed of small compact loops~\citep{golub74, kwon10}, it will be sufficient to numerically simulate a small loop system to model coronal BPs. It is well known that due to low plasma beta of the lower corona, plasma is confined along the coronal magnetic field lines. This enforces the plasma to show one dimensional (1D) dynamics along the field lines, allowing us to model the coronal loop in 1D. While observing through space or ground based telescopes, however, it is not possible to resolve every individual strand of an observed coronal loop. Rather, what we observe as a global loop is actually an ensemble of several 1D strands together. It is universally accepted that swarms of tiny reconnections called nanoflares~\citep{parker88}, each releasing energy around $10^{23}$~ergs, are responsible for the high temperature (above $1$~MK) of coronal loops.

The purpose of this work is to see if such nanoflare like heating events can give rise to density fluctuations that are reflected in the high density chromosphere before bouncing back to the corona again, thereby emulating the observed oscillations in the coronal BPs~\citep{abhishek11,chandra13}. To that end, we employ the 1D hydrodynamic coronal loop code developed in~\citet{sarkar08}. To replicate the multi-stranded nature of the global coronal loop, we simulate 125 1D hydrodynamic strands individually. These simulated strands are heated by nanoflare like heating events. Since BP loops are very tiny ($\sim10$~Mm, see below) compared to active region loops, we have assumed that loop expansion~\citep{mikic13} or thermal non-equilibrium~\citep{lionello13} does not play a crucial role in the dynamics of the loops. From the simulated global loop, we produce observables like intensity, and hence deduce the oscillation periods to see if they agree with the real oscillations studied through various instruments~\citep{abhishek11,chandra13}.

The rest of the paper is organized as follows. In Section~\ref{sec:looplength} we briefly discuss the observations of~\cite{chandra13} and from the observed BPs we find out the typical loop length which we take as input of our simulation. In Section~\ref{sec:simulation} we describe the simulation details of this multithreaded coronal loop model. In Section~\ref{forward_model} we describe the forward modeling to produce loop intensity in the AIA bands, which we analyze in Section~\ref{analysis}. Finally we conclude in Section~\ref{sec:summary} with a summary of this work and proposal for future observations.
\section{Observed intensity fluctuation and loop length determination}\label{sec:looplength}
In~\citet{chandra13}, ten different BPs were observed using full-disk Level~1.0 AIA and HMI images obtained over a period ranging from February 13 to February 15, 2011. It was found that each BP intensity varies with time. Wavelet analysis on every BP image was also performed to measure the prominent periodicities. Further details of this work can be found in section~2.1 of~\citet{chandra13}. In the present work, we want to compare BP intensity periodicities computed from our simulations with those measured by~\citet{chandra13}.

\begin{figure}
	\includegraphics[width=0.5 \textwidth,angle=-90]{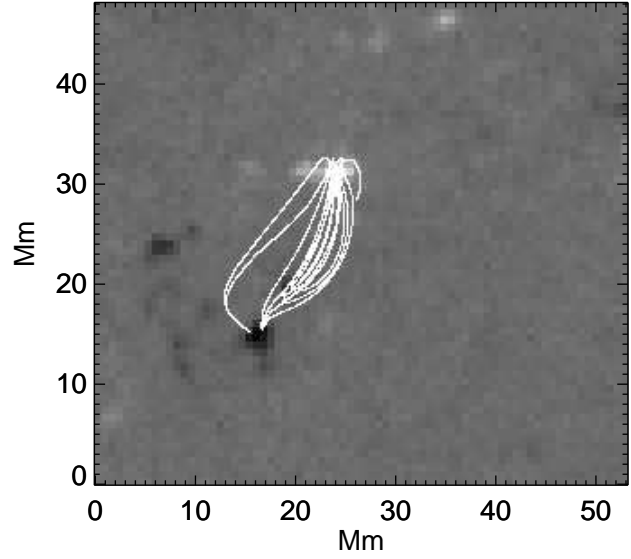}
	\caption{Graphical depiction of potential field extrapolation for BP7.}
	\label{bp7_extrapolation}
	\end{figure}

\begin{figure}
	\includegraphics[width=0.5\textwidth,angle=-90.]{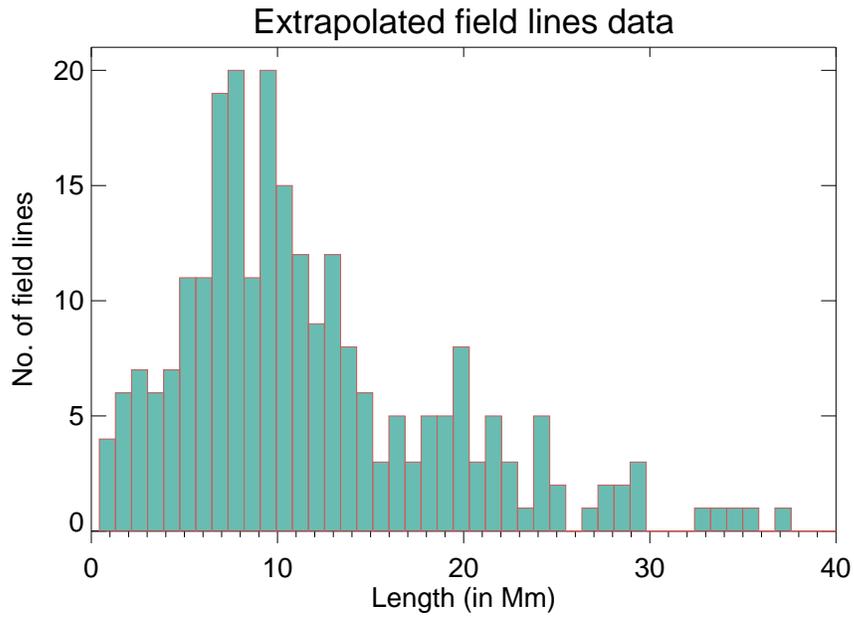}
	\caption{Histogram of the extrapolated field lines for the BPs.}
	\label{hmi_extrapolated_lengths}
	\end{figure}

As mentioned earlier, we plan to simulate a multistranded small coronal loop which can represent a typically observed BP. Thus, we need to know the typical loop length of the BP system observed by~\cite{chandra13}, which can then be used as an input parameter in the loop simulation. To that end, we make use of the HMI images corresponding to different BPs studied by~\citet{chandra13} to produce the potential field lines of force-free magnetic fields~\citep{1981A&A...100..197A,1972SoPh...25..127N}. A graphical depiction of this potential field extrapolation for a typical BP is shown in Fig~\ref{bp7_extrapolation}. Here, white lines are the potential field lines drawn using a constant $\alpha$ force-free magnetic field (with $\alpha = 0$) connecting different polarities of BP7 (fig. 1 of~\cite{chandra13}). As these lines are very short and closes up sufficiently below the source surface, it is safe to use the potential field extrapolation. We calculated lengths of different field lines composing BP7. The procedure is repeated for BP2, BP3, BP5, BP8, and BP10. BPs which had their opposite polarities very nearby were left out of this calculation since their loop lengths cannot be considered as typical. The resulting histogram of the lengths of the field lines is shown in Fig.~\ref{hmi_extrapolated_lengths}. We observe that the histogram of lengths peaks around $10$~Mm. This length have been used in our subsequent multithreaded nanoflare heated hydrodynamic loop simulation.
\section{Numerical simulation of the multistranded loop}\label{sec:simulation}
The present simulation was performed with a 1D hydrodynamical code based on Lagrange-Remap method~\citep{arber01}. In this simulation, we constructed a global loop using 125 strands. Each strand was evolved hydrodynamically independently of each other by solving the mass balance, momentum balance and energy balance equations in 1D given as follows:
	\begin{equation}\label{eq:massbalance}
	\frac{D \rho}{D t} + \rho \frac{\partial v}{\partial s} = 0~,
	\end{equation}
	\begin{equation}\label{eq:momentumbalance}
	\rho \frac{D v}{Dt} = -\frac{\partial p}{\partial s} + \rho g \cos\theta +\rho \nu \frac{\partial^2 v}{\partial s^2}~,
	\end{equation}
	\begin{equation}\label{eq:energybalance}
	\frac{\rho^{\gamma}}{\gamma - 1}\frac{D}{Dt}\bigg(\frac{p}{\rho^{\gamma}}\bigg)=\frac{\partial}{\partial s}\bigg( \kappa \frac{\partial T}{\partial s}\bigg)-n^2 Q(T) + H(s,t)~,
	\end{equation}
	\begin{equation}\label{eq:gaslaw}
	p = \frac{R}{\tilde{\mu}}\rho T~.
	\end{equation}
Here $t$ is time, $s$ represents the coordinate along the strand which is assumed to be semicircular, $\rho$, $v$, $p$, $n$ and $T$ are mass density, bulk velocity, pressure, particle density and temperature of the plasma respectively, and
	\begin{equation*}\label{eq:lagrangian}
	\frac{D}{Dt} \equiv \frac{\partial}{\partial t}+v\frac{\partial}{\partial s}~.
	\end{equation*}
We require $-L \leq s \leq L$ with $L = 5$~Mm, so that total length of the loop is $10$~Mm.~In equation~\eqref{eq:momentumbalance}, $g\cos\theta$ represents the projected component of the solar surface gravitational acceleration along the semicircular loop, $g$ being the magnitude of the acceleration. Because we are considering a small loop, we assume $g$ to be a constant with a value equal to the surface value of $2.74 \times 10^4$~cm s$^{-2}$. 
~Also in the same equation, the coefficient of viscosity $\boldsymbol{\nu = 2.0 \times 10^{14}}$~cm$^2$\,s$^{-1}$ is assumed to be uniform throughout the plasma. In equation~\eqref{eq:energybalance}, $\gamma = 5/3$ is the adiabatic index of the medium, while \mbox{$\kappa = 9.2 \times 10^{-7}T^{5/3}$~erg s$^{-1}$cm$^{-1}$K$^{-1}$} is the thermal conductivity of the plasma along the loop. Furthermore, the plasma is also cooled through radiation which is modeled by an optically thin radiative loss function $Q(T)$ based on~\cite{rosner78}. As well, $H(s,t)$ represents the heating function (whose characteristics are described in more details below) and emulates nanoflare like heating events. Finally in equation~\eqref{eq:gaslaw}, $R = 8.3 \times 10^7$ erg mol$^{-1}$K$^{-1}$ is the molecular gas constant and $\tilde{\mu} = 0.6$ mol$^{-1}$ is the mean molecular weight of the plasma.

In the present simulation, the initial and boundary conditions imposed on a strand are explained in equations (6)-(8) of~\citet{sarkar08}. These conditions emulate a strand with zero initial velocity and an initial temperature of $10^4$~K. The pressure and density initially maintain exponential profiles to represent gravitationally stratified plasma.

As already mentioned, the individual strands are heated due to artificial nanoflare like heating events. These nanoflares can occur only in the coronal part of the loop. In the simulation, we chose the location of these nanoflares randomly using pseudo random number generator. Each such event lasted from $50$s to $150$s, containing energy around $10^{23}$~ergs. The effect of a sample nanoflare containing heat $1.049 \times 10^{24}$ erg on such a strand was tested in~\cite{sarkar08} and the result was demonstrated in Figures 1 and 2 of the same article.

Due to lack of observational statistics, frequency distribution of nanoflares are still not well determined. If $f$ is the frequency distribution and $E$ is the energy, the distribution can be represented by $df/dE = E_0 E^{-\beta}$, where $E_0$ is a constant and $\beta$ is called the power-law index. In case of flares, where sufficient data are available the value of $\beta$ is equal to $1.8$. On the other hand,~\cite{hudson91} pointed out that to keep the corona heated predominantly by nanoflares the energy distribution slope has to be steeper ($\beta > 2$). Recent observation of picoflares, with estimated energy $\sim 10^{21}$ erg, also suggests the power law index $\beta$ to be in between $2.2$ to $2.7$. In the following simulations, we will primarily use $\beta = 2.3$. However, \cite{sarkar08} has demonstrated the effect of changing $\beta$ on the global loop temperature, and in the present work we also verified how changing the value of $\beta$ can affect the overall outcome.

\begin{figure}
	\includegraphics[width=0.6\textwidth,angle=-90]{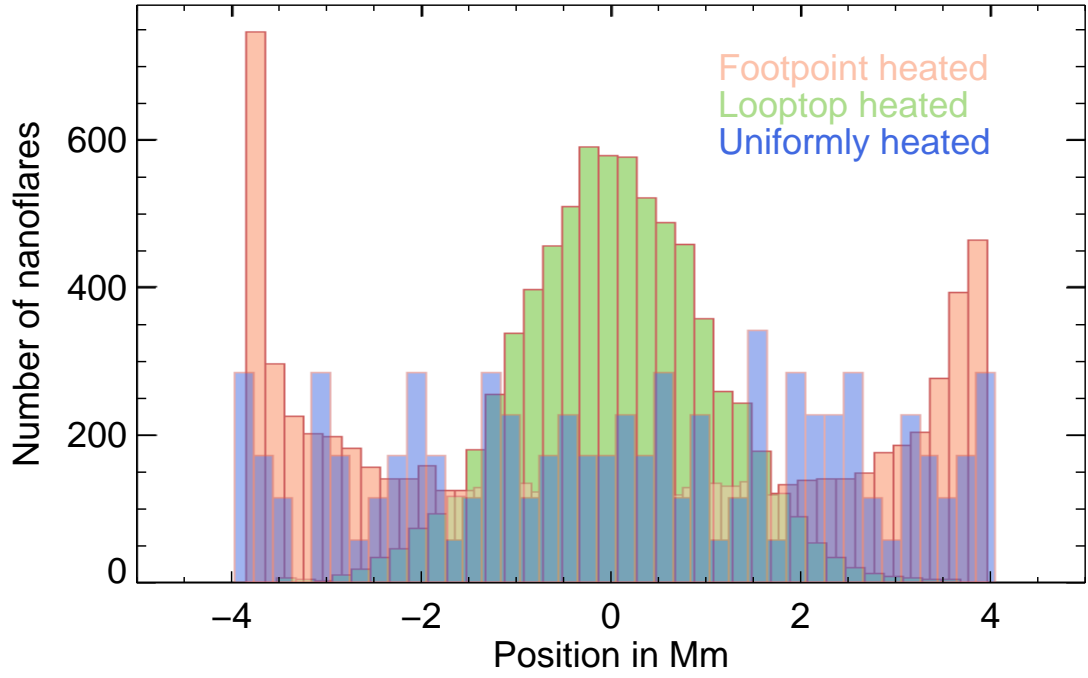}
	\caption{Histogram showing preferential heating for three different cases: footpoint, looptop and uniformly heated loops.}
	\label{heating_locations}
	\end{figure}

The temperature profile of a loop is known to be affected depending on whether it is footpoint heated, looptop heated or uniformly heated~\citep{sarkar09}. For a uniformly heated loop nanoflares can occur anywhere along the loop. For a footpoint heated loop more number of nanoflares are generated close to the footpoints. Finally, for looptop heated loops more weight is given to the looptop. A location histogram of the above three cases are shown in Fig.~\ref{heating_locations}. Part of our study, as described below, was dedicated to understand the dependence of the oscillation periods of the BP intensities on these location preferential heating.

The energy input rate per unit area per unit time would also change the loop temperature, thereby potentially affecting the intensity oscillation periods of the BPs. In this work, we have considered two cases of fixed heating rates, one at $\mathbf{4 \times 10^5}$~ergs~cm$^{-2}$~sec$^{-1}$~\citep{sarkar08} and another at $\mathbf{8 \times 10^5}$~ergs cm$^{-2}$ sec$^{-1}$.

The data synthesized through the above simulation next needs to be processed to produce observables (like intensities) to correspond to measurements of the AIA. We describe this in the following section.
\section{Production of observables}\label{forward_model}
From the simulated data, the calculated temperature and density of a strand was folded through the response function of three AIA passbands, namely 171, 193 and 211 to produce the synthetic emission. This gives the intensity $I_{\lambda, i}$ of the strand $i$ at passband $\lambda$ as follows
	\begin{equation}
	I_{\lambda,i} = G_{\lambda}(T)n_i^2(s,t)ds~,
	\end{equation}
where $G_{\lambda}(T)$ is the temperature response function at passband $\lambda$, $n_i$ is the density of the $i^\textrm{th}$ strand which is a function of space ($s$) and time ($t$), and $ds$ is a line element along the strand. One can then derive the overall intensity $I_{\lambda}$ of the global loop at passband $\lambda$ by summing over the emission from all the individual strands as follows
	\begin{equation}\label{eq:Il}
	I_{\lambda} = \sum_{i=1}^{125}I_{\lambda,i}~.
	\end{equation}
In order to match with the observation of~\citet{chandra13}, we next needed to degrade the resolution of the computed intensity~\eqref{eq:Il}.
~For a meaningful comparison of the synthetic intensity fluctuation with that observed by~\citet{chandra13}, we performed a wavelet analysis on the aforementioned synthetic data following~\citet{chandra13}. 

Like in the observation~\citep{chandra13}, we considered an eighty (80) minute time windows of the synthetic intensity from every simulation we have performed. We then convolved the time sequence of the intensity with the Morlet function to perform the wavelet analysis~\citep{1998BAMS...79...61T}. Periodicities were calculated by removing the background trend by subtracting 30 minute equivalent number of points from the intensity. The resulting periodicities are tabulated in Table~\ref{periods_table}.
                                  
\section{Analysis}\label{analysis}
\begin{figure}
	\includegraphics[width=0.7\textwidth,angle=90]{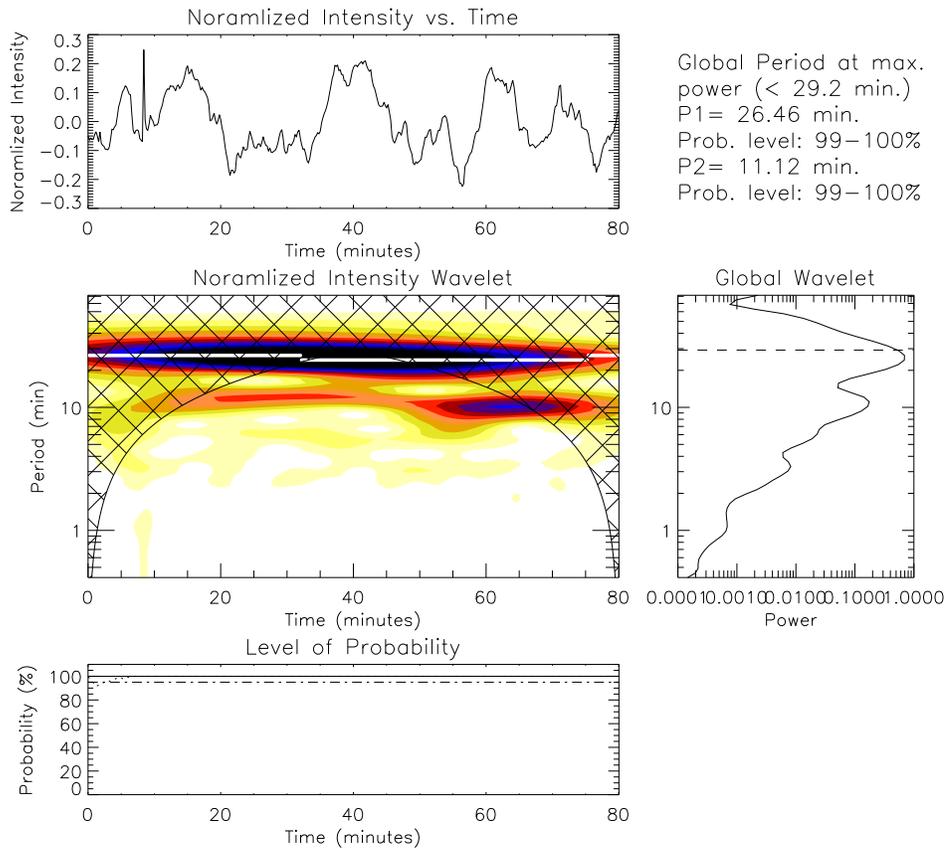}
	\caption{Wavelet analysis of the synthetic intensity (AIA 171) of a footpoint heated loop when the heat input rate is $4 \times 10^{27}$ ergs cm$^{-2}$ sec$^{-1}$. The analysis reveals P1 = $26.26$~mins and P2 = $11.12$~mins.}
	\label{wa_example}
	\end{figure}

Fig.~\ref{wa_example} showcases the results of a wavelet analysis of a typical synthetic loop intensity. The top panel shows the intensity time series at the looptop. In the middle left panel we show the wavelet power spectrum of the same intensity time series. The cross-hatched portion of this panel corresponds to locations where estimates of the oscillation period become unreliable. This region is known as the cone-of-influence (COI) according to~\citet{1998BAMS...79...61T}. The location of the maximum power is indicated by the white line in this plot. Next, the maximum measurable period is shown by a dashed line in the global wavelet spectrum plotted in the middle right panel. Finally, the bottom panel shows the variation of the probability estimate calculated using the randomization technique associated with the maximum power at each time in the wavelet power spectrum. The oscillations with a probability greater than 95\% are considered to be significant. The periods of the peak power are printed above the global wavelet plot. Here, P1 refers  to the most significant peak and P2 refers to the second most significant peak with confidence level higher than 95\%. The result shows that the power is more for the period P1 $= 26.26$~mins oscillation, where as the second peak is at P2 $= 11.12$~mins.

	\begin{figure}
	\includegraphics[width=0.9\textwidth,angle=90]{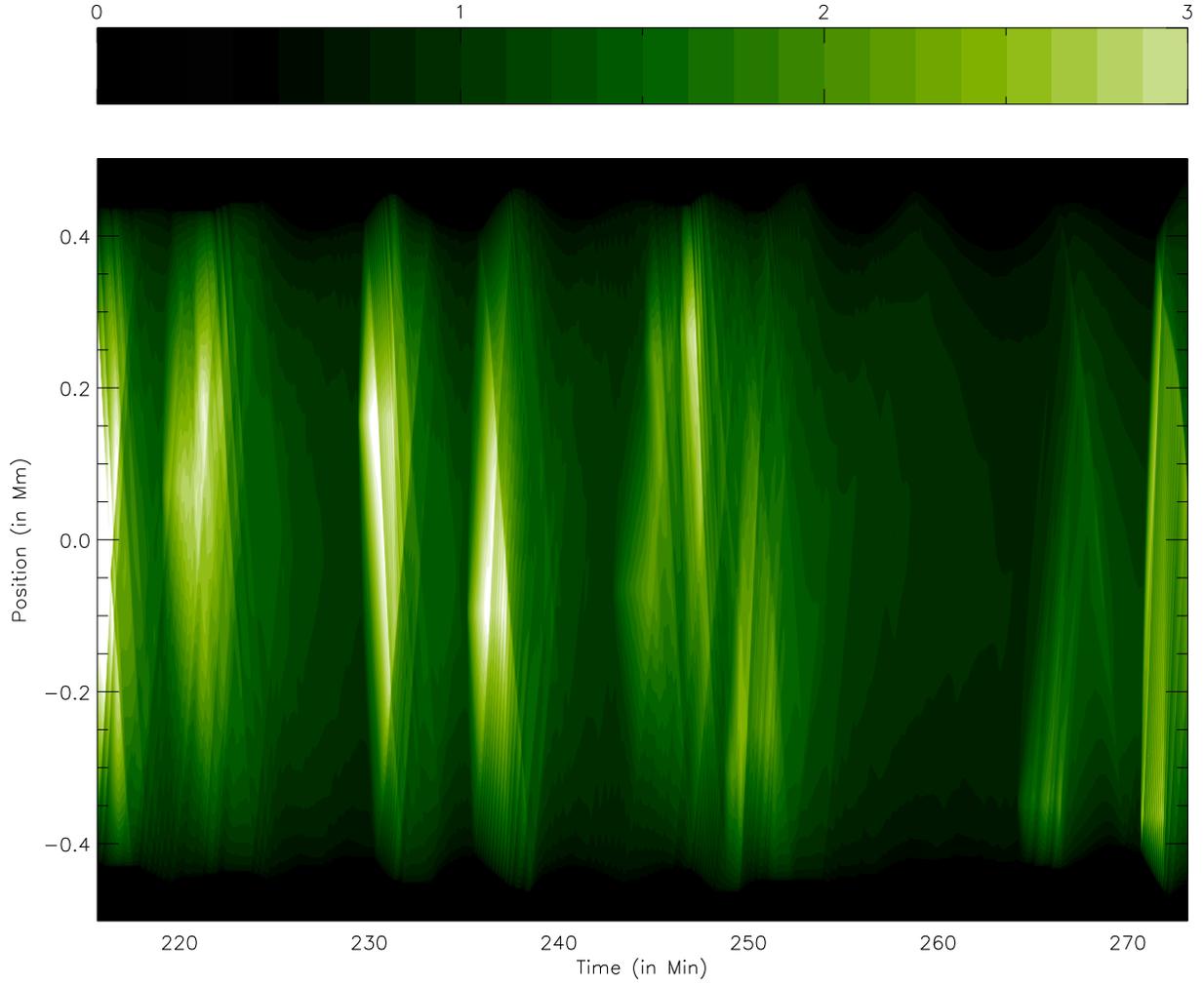}
	\caption{Evolution of $p\rho^{-\gamma}$ (in appropriate units) along the strand from the simulation of a uniformly heated loop with total heat $4 \times 10^{27}$ ergs cm$^{-2}$ sec$^{-1}$.}
	\label{adiabatic}
	\end{figure}
	\begin{figure}
	\includegraphics[width=0.9\textwidth,angle=90]{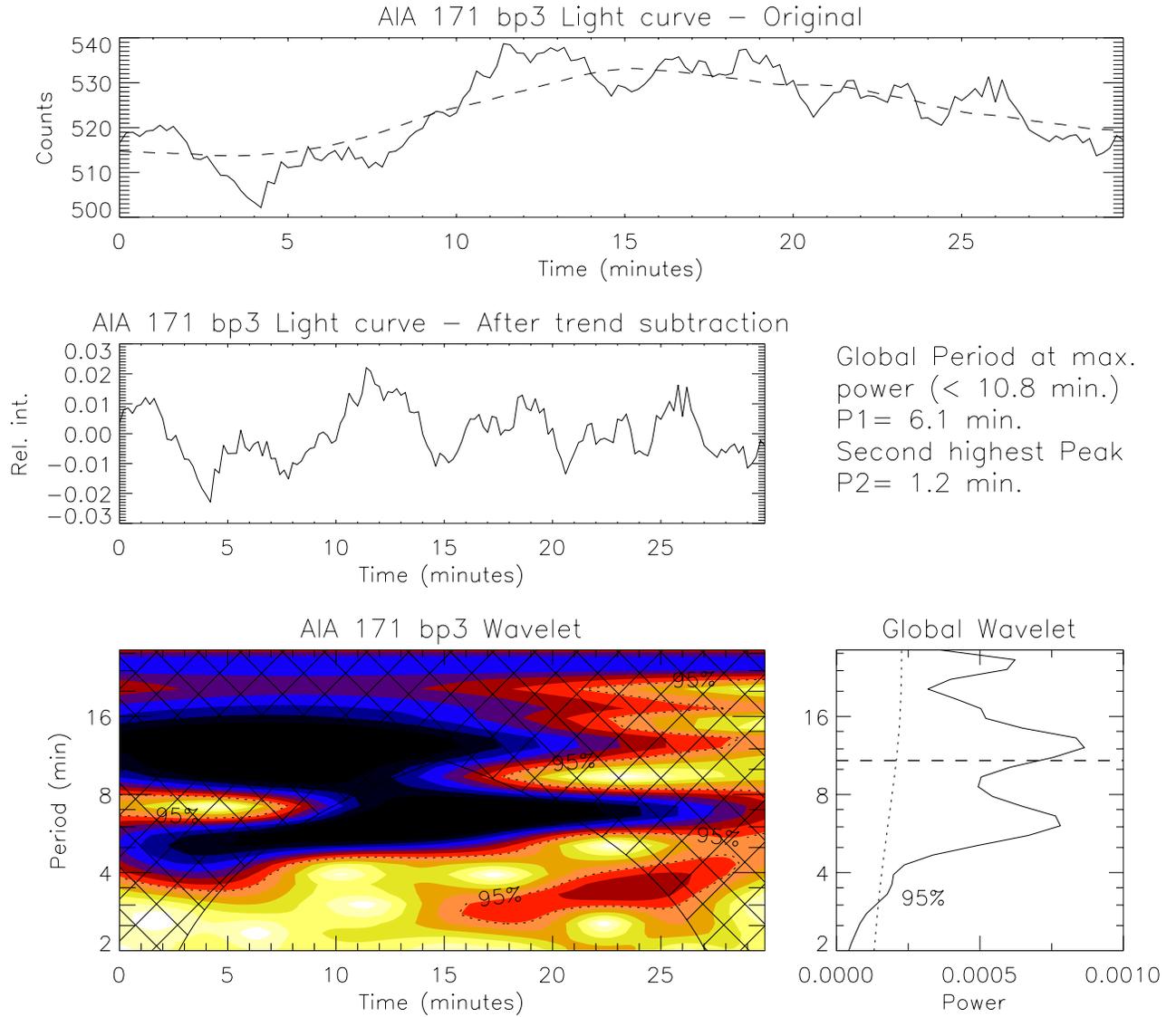}
	\caption{Wavelet analysis of BP3 intensity (171A) with $9$~mins running average subtraction, revealing periods of the fundamental mode of the standing waves.}
	\label{two_min}
	\end{figure}
	\begin{figure}
	\includegraphics[width=0.9\textwidth]{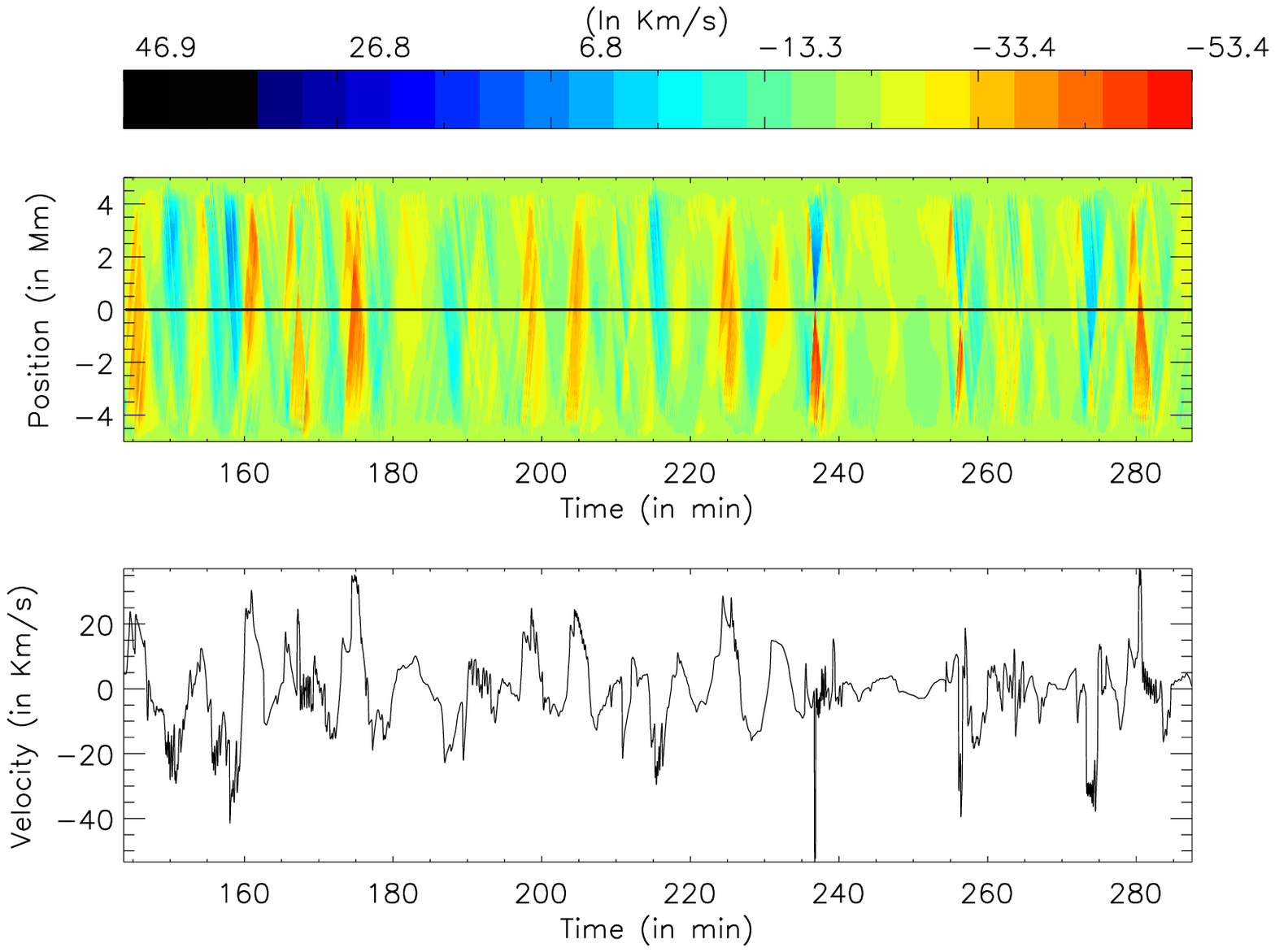}
	\caption{Uniformly heated loop strand velocity: (top) contour plot of the strand velocity evolution, (bottom) velocity evolution at the center of the loop, marked by the dark line in the top figure.}
	\label{strand_velocity}
	\end{figure}

The overall simulation results are summarized in Table~\ref{periods_table}. There appears to be no distinct difference between observed oscillation periods \citep{chandra13} and the oscillation periods from the synthetic data of $10$ Mm loops. Both the simulation and observation show typical oscillation periods of around $15$ - $25$~mins. The characteristic oscillation period of the simulated loops are also seemingly independent of the heating location or the heat input.

To obtain further insight into the origin of such intensity oscillations, it is instructive to analyze individual strand behavior. We randomly selected a strand for this purpose, e.g. strand no. 12 of the footpoint heated loop. Since intensity of this strand is proportional to the square of its density, we performed a wavelet analysis of its density squared exactly in the same way as have been done in the case of the global loop intensity. These results are also tabulated in Table~\ref{periods_table} and we find similar range for the periods ($15$ - $25$~min) as that of the intensity of the global loop.

In the past,~\citet{wang02} have observed oscillations of periods ranging from $14$ - $18$ minutes in flaring loops. It was thought that this kind of loop oscillations were excited by a single flarelike impulsive heating event. Using 1D loop simulations,~\cite{valeri04, tsiklauri04, selwa05} have later confirmed that indeed such oscillations with such periodicities can be generated using impulsive heating events. These studies have further concluded that these oscillations are due to slow standing waves in the loop.

Compared to flaring loops, BP loops are significantly smaller in length, with the majority of them having length around $10$~Mm as reported earlier in Section~\ref{sec:looplength}. The slow wave oscillation period $P$ of such waves can be calculated using
	\begin{equation}
	P = \frac{2L}{n C_s}~,
	\label{sound_speed:1}
	\end{equation}
where $L$ is the loop length, $C_s$ is the sound speed of the loop plasma, and the integer $n$ labels the harmonics. $C_s$ is given by~\citep{markus04} 
	\begin{equation}
	C_s = \sqrt{\frac{\gamma p}{\rho}} = 147\sqrt{\frac{T}{1\textrm{MK}}}~\text{[Km/s]}~.
	\label{sound_speed:2}
	\end{equation}


For our simulated strands with $L = 10$~Mm and temperature $T$ varying between $0.2$~MK to $2$~MK, the fundamental period ($n = 1$), according to equations~\eqref{sound_speed:1} and~\eqref{sound_speed:2}, should lie within the range of $5$ to $2$ mins. On the other hand, a direct analysis of the synthetic data (as shown in Table~\ref{periods_table}) as well as the observations of~\citet{chandra13} yield an order of magnitude higher periodicity (in the range of $15$ - $25$~mins) for the modes corresponding to the two most significant power peaks. Clearly, the speeds of these higher periodicity modes are not described by equation~\eqref{sound_speed:2}.~Indeed, equation~\eqref{sound_speed:2} provides the expression for sound speed of a mode which respects adiabaticity \citep{landau}. But the right hand side of the energy equation~\eqref{eq:energybalance} shows a strict violation of adiabaticity over the period of the simulation, and one may identify the nano-flares contributing to $H(s, t)$ being the primary source of this. This may be confirmed in the plot of Fig.~\ref{adiabatic} which captures a variation of $p\rho^{-\gamma}$ along the loop during a part of its evolution.

Nevertheless, modes with characteristic time scales comparable with the time scale associated with the nanoflares should approximately respect adiabaticity. Given that the nanoflare lifetime varies between $50$ - $150$ secs, one may then identify the modes with periodicity between $2$ - $5$ mins with such adiabatic modes.~To uncover these modes from the synthetic data, an analysis was performed on the synthetic intensity (193~\AA) time series data with $9$ mins background subtraction. One may contrast this with the previous wavelet analysis of the intensity time series data where a $30$ minutes equivalent number of background points were subtracted. For the present analysis, it was found $P1 = 3.03$ mins and $P2 = 2.14$ mins in case of footpoint heated loop when the heat input is $4 \times 10^{27}$ ergs cm$^{-2}$ s$^{-1}$.

We next performed a similar wavelet analysis on the data of~\citet{chandra13} to compare with the smaller background subtraction on the synthetic data. In particular, we selected $20$ mins intensity data of BP3 for this purpose. Instead of $30$ mins background subtraction, a $9$ mins background subtraction was performed during the wavelet transform. The result is shown in the Fig.~\ref{two_min}. We find through this analysis that the periodicities of the BP are $P1 = 6.1$ mins and $P2 = 1.2$ mins, which are in the same range of the periodicities obtained from the analyzed synthetic data. This strongly indicates that the oscillations in the range of $15$ - $25$ mins are due to non-adiabatic modes, which either gets generated by the nanoflares displacing the local plasma, or through the chromospheric evaporation~\citep{doschek80,feldman80} due to which the plasma gets into the coronal part of the loop.

Since the oscillations in the range of $15$ - $25$ mins, observed by~\citet{chandra13} and also captured in the present simulation, are of non-adiabatic origin, one cannot use equation~\eqref{sound_speed:2} to determine their speed. However, if one assumes that these are fundamental modes, then the speed of such modes turn out to be about $17$ Km sec$^{-1}$ according to equation~\eqref{sound_speed:1}. Interestingly, an analysis of the general plasma flow for a typical strand (strand 13) of the uniformly heated loop where the heat input rate is $4 \times 10^{27}$ ergs cm$^{-2}$ s$^{-1}$ reveals a velocity time evolution along the strand as shown in Fig.~\ref{strand_velocity}. According to this plot, the plasma velocity at the center of the strand can reach up to $20$ Km sec$^{-1}$, which is very close to the estimated speed of the modes presented above. Therefore, its seems highly plausible that the oscillations in the range of $15$ - $25$ mins are born as a result of hydrodynamic shock waves generated by the nanoflares.
\section{Summary and Discussion}\label{sec:summary}
Coronal BPs have been observed since long~\citep{vaiana73} and are known to be small loop systems. It has also been known that the intensity of these BPs oscillate within certain range of periodicities~\citep{tian08, abhishek11, chandra13}. However the origin of these oscillations is still unknown. In this work, we propose a mechanism behind such oscillation by modeling the BPs as strands of plasma heated by swarms of nanoflares.

The typical loop length of such BP loops were computed using potential field extrapolation (Fig.~\ref{bp7_extrapolation}) and was found to be around $10$ Mm.~These BP loops were then modeled using a 1D multithreaded numerical code~\citep{sarkar08}, with the length set to $10$ Mm. Strands of these simulated loops were heated by series of nanoflare like heating events during their lifetime. We considered three kinds of loops depending on the heating location, namely footpoint heated, looptop heated, and uniformly heated.

The oscillation periods and temperature of the simulated loops are given in the Table~\ref{periods_table}. We observed that differences in the heating location do not affect the oscillation periods.  We also changed the heat input rate from $4 \times 10^{27}$ erg cm$^{-2}$s$^{-1}$ to $8 \times 10^{27}$ erg cm$^{-2}$s$^{-1}$ in order to study the effect of increase of temperature of the loop on the periodicity. However, no significant correlation between the loop temperature and the oscillation period was observed.
It is also worth checking if the nanoflare distribution function has any effect on the oscillation period. For that we ran a simulation of the uniformly heated global loop (heat input rate $4 \times 10^{27}$~erg cm$^{-2}$s$^{-1}$) with power law index $\beta=3.29$. The wavelet analysis of the loop intensity (AIA 193) yields $P1=18.71$~mins and $P2=9.35$~mins, when $30$~mins background substraction was performed.
This suggests the observed oscillation period is unaltered with the change of power law index $\beta$.

Table~\ref{periods_table} shows that the oscillation periods of the simulated loops were in the range of $15$ - $25$ mins and such range compares well with the values observed by~\citet{chandra13}. These oscillations are present both in the intensity time series of the global loop as well as in the fluctuations of the individual strand density. It is suggested that these oscillations are due to non-adiabatic shock waves present in such multiple nanoflare heated loops, since their periods are much higher than those for adiabatic modes generated for the temperature range $0.2$ - $2.0$ MK of the loops, and also because the speed of these modes are in the same range as the typical speed of the plasma ($\sim 20$~Km sec$^{-1}$). In fact, for the given temperature range of the plasma, the adiabatic modes have periodicities in the range $2$ - $5$ mins according to equation~\eqref{sound_speed:2}. These adiabatic modes can be uncovered both from the simulated data as well as from the observations~(\citet{chandra13}, from the intensity light curve of BP3, a typical BP) through a $3$ mins background subtraction instead of a $30$ mins background subtraction as originally performed by~\citet{chandra13}. 

To summarize, in this work we have proposed two kinds of oscillations in coronal BP intensities, namely oscillations due to shock waves and oscillations due to adiabatic modes. Further confirmation of this proposal can be achieved by studying the Doppler velocity map or non-thermal line broadening of BPs. These velocity maps and line broadenings can then be compared with the simulated synthetic data resulting in a better understanding of such waves.

We are grateful to the anonymous referee for carefully reading the manuscript and suggesting us important changes. The Center of Excellence in Space Science India (CESSI) is supported by the Ministry of Human Resource Development, Government of India.

\begin{landscape}

\begin{table}[h]
\caption{Periodicities observed in simulated loops.\label{periods_table}}
\label{my-label}
\begin{tabular}{|c|c|c|c|c|c|c|c|c|c|c|}
\hline
\multicolumn{1}{|l|}{\multirow{4}{*}{}}                                     & \multirow{4}{*}{\begin{tabular}[c]{@{}c@{}}Energy input\\ \\ \\ ($ergs/cm^{2}/sec$)\end{tabular}} & \multicolumn{6}{c|}{Intensity}                                                                                                                                                                                                                                                                                                                                                                                                          & \multicolumn{2}{c|}{\multirow{2}{*}{\begin{tabular}[c]{@{}c@{}}Density squared\\ (single strand)\end{tabular}}}                             & \multirow{2}{*}{\begin{tabular}[c]{@{}c@{}}Looptop temperature\\ (MK)\\ \end{tabular}} \\ \cline{3-8}
\multicolumn{1}{|l|}{}                                                      &                                                                                                  & \multicolumn{2}{c|}{171}                                                                                                                    & \multicolumn{2}{c|}{193}                                                                                                                    & \multicolumn{2}{c|}{211}                                                                                                                    & \multicolumn{2}{c|}{}                                                                                                                       &                                                                                        \\ \cline{3-11} 
\multicolumn{1}{|l|}{}                                                      &                                                                                                  & \multirow{2}{*}{\begin{tabular}[c]{@{}c@{}}P1\\ (mins)\end{tabular}} & \multirow{2}{*}{\begin{tabular}[c]{@{}c@{}}P2\\ (mins)\end{tabular}} & \multirow{2}{*}{\begin{tabular}[c]{@{}c@{}}P1\\ (mins)\end{tabular}} & \multirow{2}{*}{\begin{tabular}[c]{@{}c@{}}P2\\ (mins)\end{tabular}} & \multirow{2}{*}{\begin{tabular}[c]{@{}c@{}}P1\\ (mins)\end{tabular}} & \multirow{2}{*}{\begin{tabular}[c]{@{}c@{}}P2\\ (mins)\end{tabular}} & \multirow{2}{*}{\begin{tabular}[c]{@{}c@{}}P1\\ (mins)\end{tabular}} & \multirow{2}{*}{\begin{tabular}[c]{@{}c@{}}P2\\ (mins)\end{tabular}} & \multicolumn{1}{l|}{\multirow{2}{*}{}}                                                 \\
\multicolumn{1}{|l|}{}                                                      &                                                                                                  &                                                                      &                                                                      &                                                                      &                                                                      &                                                                      &                                                                      &                                                                      &                                                                      & \multicolumn{1}{l|}{}                                                                  \\ \hline
\multirow{2}{*}{\begin{tabular}[c]{@{}c@{}}Looptop\\ heated\end{tabular}}   & $4 \times 10^{27}$                                                                                          & 15.7                                                                 & 24.3                                                                 & 17.15                                                                & 10.2                                                                 & 17.2                                                                 & 9.35                                                                 & 22.3                                                                 & 8.6                                                                  & 2.5                                                                                    \\ \cline{2-11} 
                                                                            & $8 \times 10^{27}$                                                                                          & --                                                                   & --                                                                  & 22.3                                                                 & 11.1                                                                 & 20.4                                                                 & 11.1                                                                 & 25.3                                                                 & 11.1                                                                 & 3.1                                                                                    \\ \hline
\multirow{2}{*}{\begin{tabular}[c]{@{}c@{}}Footpoint\\ heated\end{tabular}} & $4 \times 10^{27}$                                                                                          & 26.3                                                                 & 11.1                                                                 & 11.1                                                                 & 2.6                                                                  & 11.1                                                                 & 26.5                                                                 & 17.1                                                                 & 13.2                                                                 & 2.2                                                                                    \\ \cline{2-11} 
                                                                            & $8 \times 10^{27}$                                                                                          & --                                                                   & --                                                                   & 22.3                                                                 & 7.2                                                                  & 20.4                                                                 & 7.2                                                                  & 14.4                                                                 & 7.9                                                                  & 2.7                                                                                    \\ \hline
\begin{tabular}[c]{@{}c@{}}Uniformly\\ heated\end{tabular}                  & $4 \times 10^{27}$                                                                                          & 17.2                                                                 & 26.5                                                                 & 17.2                                                                 & 7.9                                                                  & --                                                                   & --                                                                   & 18.7                                                                 & 8.6                                                                  & 2.2                                                                                    \\ \hline
\end{tabular}
\end{table}

\end{landscape}

\bibliography{mybib}{}
\end{document}